\documentclass[trackchanges]{aastex701}

\usepackage{epsfig}
\usepackage{alltt,color}
\usepackage{graphicx}
\usepackage{epstopdf}
\usepackage{float}
\usepackage{palatino}
\usepackage{amsmath}
\usepackage{cases}
\usepackage{pifont}
\usepackage{txfonts}
\usepackage{multirow}
\usepackage{appendix}

\begin{document}

\title{A novel mechanism for coherent radio emission: particle-in-cell simulations} 

\correspondingauthor{Xiaowei Zhou}
\email{zhouxw@pmo.ac.cn}

\author[orcid=0000-0001-7855-2479, gname=Xiaowei, sname=Zhou]{Xiaowei Zhou}
\affiliation{Purple Mountain Observatory, Chinese Academy of Sciences, Nanjing 210034, People's Republic of China}
\email[show]{zhouxw@pmo.ac.cn}


\author[orcid=0000-0003-2418-5508, gname=Dejin, sname=Wu]{Dejin Wu}
\affiliation{Purple Mountain Observatory, Chinese Academy of Sciences, Nanjing 210034, People's Republic of China}
\email{}

\author[orcid=0000-0001-8058-2765, gname=Ling, sname=Chen]{Ling Chen}
\affiliation{Purple Mountain Observatory, Chinese Academy of Sciences, Nanjing 210034, People's Republic of China}
\email{}

\author[orcid=0000-0002-9893-4711, gname=Zongjun, sname=Ning]{Zongjun Ning}
\affiliation{Purple Mountain Observatory, Chinese Academy of Sciences, Nanjing 210034, People's Republic of China}
\email{}

%

\begin{abstract}
It has been established that electron (phase-space) holes, i.e., EHs, can arise from the nonlinear relaxation of the electron beam instability. Excitation of electromagnetic waves related to the EHs is, however, still not fully understood. 
Using 2.5-dimensional particle-in-cell simulations, we systematically investigate the nonlinear evolution of the (moderately relativistic) electron bi-stream instability (a special case of the electron beam instability where the densities of the two counter-propagating energetic electron beams are equivalent) in charge- and current-neutralized plasmas under weak and strong magnetizations. 
Our results reveal that the background magnetic field fundamentally alters the real- and phase-space structures (i.e., potential-$\phi$ wells and EHs) and subsequent wave excitations. In the weakly magnetized plasma, the perpendicular $\vec{E}_{\perp} \times \vec{B}_0$ drift, associated with the $\phi$ wells, breaks the velocity-space symmetry, generating a positive gradient ($\partial f/\partial U_\perp > 0$) and significantly enhanced perpendicular currents. These localized currents act as free-energy sources for exciting perpendicularly propagating electrostatic Bernstein and electromagnetic electron cyclotron harmonics, which ultimately result in stronger and approximately isotropic electron heating in the weakly magnetized plasma.
In contrast, the strong magnetic field suppresses electron perpendicular drifts and confines the dynamics to the parallel direction. Consequently, wave excitation is predominantly driven by the parallel current, leading to the dominance of parallel electrostatic (e.g., Langmuir) waves and obliquely propagating electromagnetic (e.g., whistler, Z-, and O-mode) waves. Furthermore, the strongly magnetized plasma facilitates much larger-amplitude and broader-spatial-extent $\phi$ wells, which allow a greater fraction of electrons near the $\phi$ well edges to be accelerated to higher velocities via wave-surfing (surfatron) acceleration, resulting in phase-space bifurcations and distinct high-velocity tails. 
These findings provide new insights into the wave excitation, the nonlinear wave-particle interactions, the formation of non-Maxwellian features, and the anisotropic electron heating in magnetized plasmas, with potential implications for coherent radio emission in astrophysical plasmas (e.g., solar zebra-pattern bursts).
\end{abstract}



\section{Introduction} 
\label{introduction}

An electron phase-space hole (EH) is a localized, self-consistent electrostatic phase-space structure in collisionless plasmas, formed by the phase-space trapping of electrons. It is characterized by: (1) a localized hole or vortex encircled by trapped electron orbits in position--velocity phase space; (2) a localized positive electrostatic potential hump in real (spatial--temporal) space, implying a local positive net charge (i.e., $n_i>n_e$) at its location and demonstrating that EHs are predominantly electrostatic; and (3) a bipolar parallel electric field ($E_{\parallel}$ along the ambient magnetic field $B_0$) signature that coincides with the positive electrostatic potential hump~\citep{Hutchinson_2017PhPl...24e5601H}. 
Note that this generalized definition of an EH relaxes the constraints of the ideal 1D steady Bernstein--Greene--Kruskal (BGK) solution to the time-independent Vlasov--Poisson equations~\citep{Bernstein_1957PhRv..108..546B} and permits non-stationary, dissipative, multidimensional, asymmetric EHs, which have been detected in PIC/Vlasov simulations and space plasma observations.

EHs have been observed in various regions of the Earth’s magnetosphere, e.g., in the magnetotail, magnetosheath, auroral region, magnetic reconnection region, magnetopause, and at the bow shock~\citep{Matsumoto_etal_1994GeoRL..21.2915M, Ergun_etal_1998GeoRL..25.2041E, Franz_etal_1998GeoRL..25.1277F, Kojima_etal_1994GeoRL..21.2919K, Kojima_etal_1997JGR...10214439K, Mozer_etal_1997PhRvL..79.1281M, Bale_etal_1998GeoRL..25.2929B, Cattell_etal_2002GeoRL..29.1065C, Pickett_etal_2004NPGeo..11..183P, Pickett_etal_2003NPGeo..10....3P} and in the solar wind~\citep{Mangeney_etal_1999AnGeo..17..307M}, as well as in laboratory experiments of, for example, laser-generated plasma~\citep{Sarri_etal_2010PhPl...17a0701S} and magnetic reconnection~\citep{Fox_etal_2008PhRvL.101y5003F}. Properties (e.g., amplitude, spatial scale, and spatial structure) of EHs depend on the regions where they are detected. 
For instance, in the plasma sheet boundary layer of the magnetotail, most of EHs observed by the Geotail spacecraft show one-dimensional (1D) bipolar $E_{\parallel}$ structures with little variation of electric fields perpendicular to the ambient magnetic field ($E_{\perp}$). The $E_{\parallel}$ amplitude of these EHs are also very small~\citep{Matsumoto_etal_1994GeoRL..21.2915M, Kojima_etal_1994GeoRL..21.2919K, Kojima_etal_1997JGR...10214439K}.
In the auroral region, however, bipolar $E_{\parallel}$ accompanied by strong $E_{\perp}$ as well as impulsive magnetic field fluctuations are observed by the Polar (e.g.,~\citealp{Mozer_etal_1997PhRvL..79.1281M}) and FAST (e.g.,~\citealp{Ergun_etal_1998GeoRL..25.2041E}) spacecrafts. This implies that large-amplitude EHs can have multi-dimensional structures and lead to electromagnetic emission. 

Verious particle-in-cell (PIC) simulations and theoretical studies have been performed to figure out the generation of EHs in the different regions of the Earth’s magnetosphere. Generally, nonlinear evolution of the electron beam instabilities (e.g., the electron/ion two-stream or Buneman instability, the electron bump-on-tail instability, and the electron two-stream instability, which are common in space and astrophysical plasma environments~\citealp{Gary_1993tspm.book.....G}) are considered to be the main generation mechanisms for EHs~\citep{Omura_etal_1994GeoRL..21.2923O, Singh_etal_2000GeoRL..27.2469S, Singh_etal_2001JGR...10621183S, Miyake_etal_1998JGR...10311841M, Lu_etal_2005JGRA..110.3223L, Morse&Nielson_1969PhFl...12.2418M, Umeda_etal_2006JGRA..11110206U, Miyake_etal_2000JGR...10523239M, Lu_etal_2008JGRA..11311219L, Omura_etal_1996JGR...101.2685O, Omura_etal_2001Ap&SS.277...45O, Umeda_etal_2004JGRA..109.2207U, Wu_etal_2012Ap&SS.338...81W, Wu_etal_2013PlST...15...17W}.
According to these PIC simulation investigetions, generally, the amplitude of EHs can be larger in the case of the two-stream instability than the bump-on-tail instability. A electrostatic potential hump led by the two-stream instability could trap the whole population of two electron beam components.  Therefore, the electron two-stream instability has been assumed to be a generation model for EHs with large amplitude~\citep{Umeda_etal_2006JGRA..11110206U}. 
Since EHs are predominantly electrostatic, most of the above mentioned PIC simulations and theoretical works were performed with electrostatic approximation. \citealp{ Umeda_etal_2004JGRA..109.2207U, Wu_etal_2013PlST...15...17W}, for example, considered the electromagnetic field signatures associated with EHs in the Earth’s magnetosphere via 2.5D electromagnetic PIC simulations. Both of these investigations confirmed that the current driven by the $\vec{E}_{\perp} \times \vec{B}_0$ drift of electrons in EHs is capable of giving rise to electromagnetic emissions. However, they did not address the properties of the excited electromagnetic waves (e.g., wave mode, emission anisotropy). 

Investigations mentioned above mostly focus on environment of the Earth’s magnetosphere. As is well known, however, energetic electron beams are not unique to Earth. Accelerated high-energy electron beams are present in a broad variety of astrophysical environments, including the Sun, magnetospheres of giant planets (e.g., Saturn), supernova remnants, pulsar wind nebulae. In these systems, electrons typically reach higher energies than those found within the Earth’s magnetosphere~\citep{Marcowith_etal_2016RPPh...79d6901M, Kahler_2007SSRv..129..359K, Kollmann_etal_2018JGRA..123.9110K}.
Formation and evolution of EHs driven by more energetic and denser electron beams, as well as their subsequent impacts on electromagnetic wave excitations, remain to be fully elucidated.

Hence, in this paper, focusing on the electron bi-stream instability (a special case of the electron beam instability where the densities of the two counter-propagating energetic electron beams are equivalent), we carry out 2.5D fully kinetic (electromagnetic) PIC simulations to investigate the following four aspects in differently magnetized plasmas: (1) the nonlinear evolution of the electrostatic potential ($\phi$) and the resulting EHs, (2) properties of the excited electrostatic and electromagnetic waves (e.g., wave modes and propagation directions), (3) the roles of these $\phi$ structures in driving wave excitations, and (4) the resulting electron acceleration and heating properties. This paper is organized as follows: Section~\ref{setup} describes the simulation model and numerical setup. Section~\ref{Results} presents the simulation results. The conclusions (Section~\ref{conclusions}) and discussion (Section~\ref{discussion}) of this study are presented at the end.

\section{Simulation Model}
\label{setup}

We perform this study with the fully kinetic PIC code --- ACRONYM~\citep{Kilian_etal_2017_PoP}, a fully relativistic electromagnetic code tuned for the study of kinetic-scale plasma wave phenomena and interactions in collisionless plasmas~\citep[see, e.g.,][] {Ganse_etal_2012ApJ...751..145G, Kempf_etal_2016A&A...585A.132K, Schreiner_etal_2017ApJ...834..161S, Munoz2018PhRvE, Zhou_etal_2020ApJ...891...92Z}. We use its version in two spatial dimensions and three dimensions in velocity, components of the electromagnetic fields (i.e., 2.5-dimensional or 2.5D) for this study. All quantities in simulations are solved in real space with CGS units.

In this study, all simulations have the same spatial and time resolution. In particular, the grid cell size is $\Delta x = \Delta y \approx 1.24 \lambda_{De}$ ($\lambda_{De}$ is the Debye length) with $N_{x} \times N_{y} = 2048 \times 2048$ grid points in the $x$--$y$ plane of the 2D simulation box. Periodic boundaries are applied in both directions for both fields and particles.
The timestep ($\Delta t$) in these simulations is determined by the inherent length and timescale requirements in a fully kinetic PIC code, i.e., the Courant--Friedrichs--Lewy (CFL) condition for the speed of light $c$.

For the (numerical) particles, two species are employed to maintain the global charge neutrality: energetic electron beams and thermal protons. The proton-to-electron mass ratio is chosen as the realistic value $m_{p}/m_{e} = 1836$. 
In the electron bi-stream frame, the energetic electron beams are divided into two components of equal density but opposite drift velocities along the ambient magnetic field, with $U_{de, 0}/c = \pm 0.47$, where $c$ is the speed of light, corresponding to $\gamma_{0} = 1.1$. 
Consequently, the whole plasma system is macroscopically charge-neutral and current-free.
The initial temperature of these two electron beams are, however, considered to be different: the forward (backward) electron beam has a theremal velocity $U_{th, ef}/c = 0.03$ ($U_{th, eb}/c = 0.015$).  And the protons have the same temperature with the backward electron beam. The above $\lambda_{De}$ is defined also in terms of $U_{th, eb}/c = 0.015$.
In total, 2300 particles per cell are homogeneously distributed over the entire (spatial) simulation domain with a constant ambient magnetic field along the x-axis, i.e., $\vec{B}_{0} = B_{0} \vec{x}$.

In this paper, we consider two plasmas (Case 1 and Case 2) with different magnetization levels. The ratios between the electron gyrofrequency and the electron plasma frequency ($\omega_{ce} / \omega_{pe}$) are equal to $0.54$ and $5.4$ for Case 1 and Case 2, respectively. $\omega_{pe}$ is fixed for all simulations and $\vec{B}_{0}$ is, hence, initialized according to the frequency ratio $\omega_{ce} / \omega_{pe}$.

All parameters presented in Section~\ref{Results} are normalized quantities. Normalizations used throughout this paper are as follows: $\omega_{norm} = \omega_{pe}$ is the normalization of frequency. Time, momentum and distance are normalized by $1/\omega_{norm}$, $c$ and $c/\omega_{norm}$, respectively. $B_{norm}$ is the normalization of the electric and magnetic field strengths and corresponds to the ambient magnetic field $\vec{B}_{0}$ for $\omega_{ce}/\omega_{pe} = 5.4$ in Case 2. Intensity is normalized by the initial total magnetic field intensity in Case 2, i.e., $\varepsilon_{norm}$. 

\section{Simulation Results}
\label{Results}

\subsection{Formation and evolution of EHs}

\begin{figure*}
\begin{center}
\includegraphics[width=1.0\textwidth] {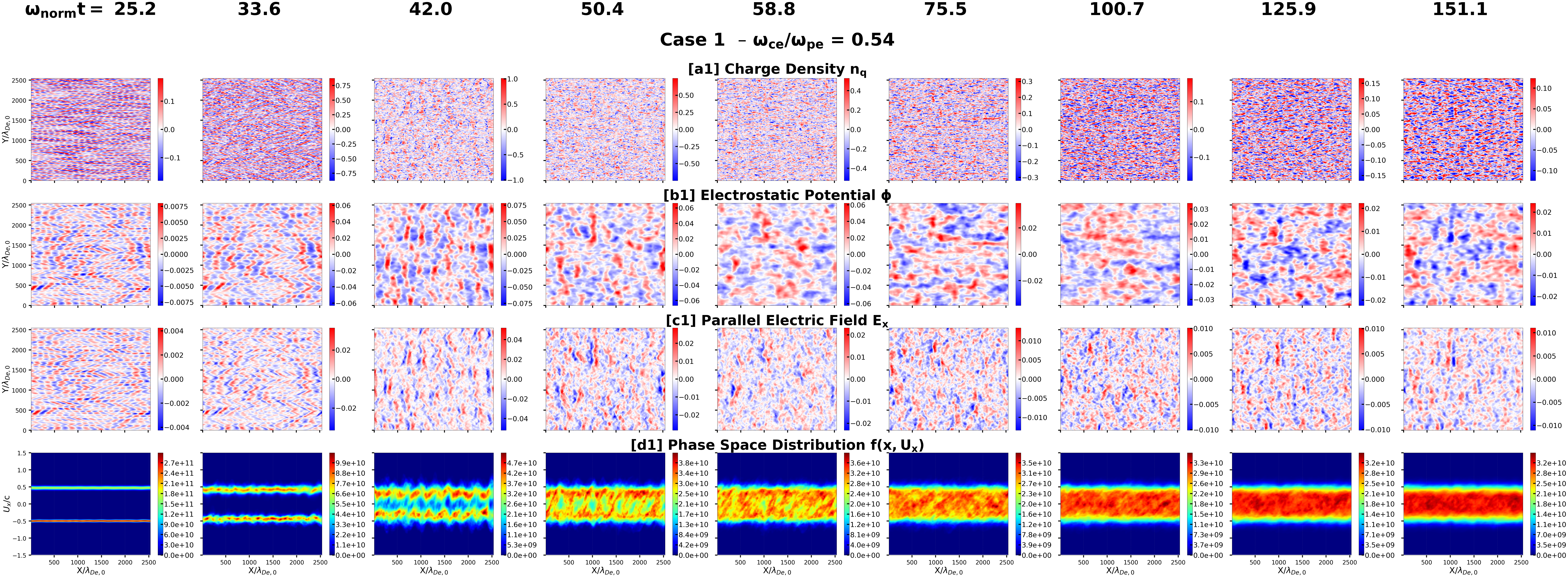}
\includegraphics[width=1.0\textwidth] {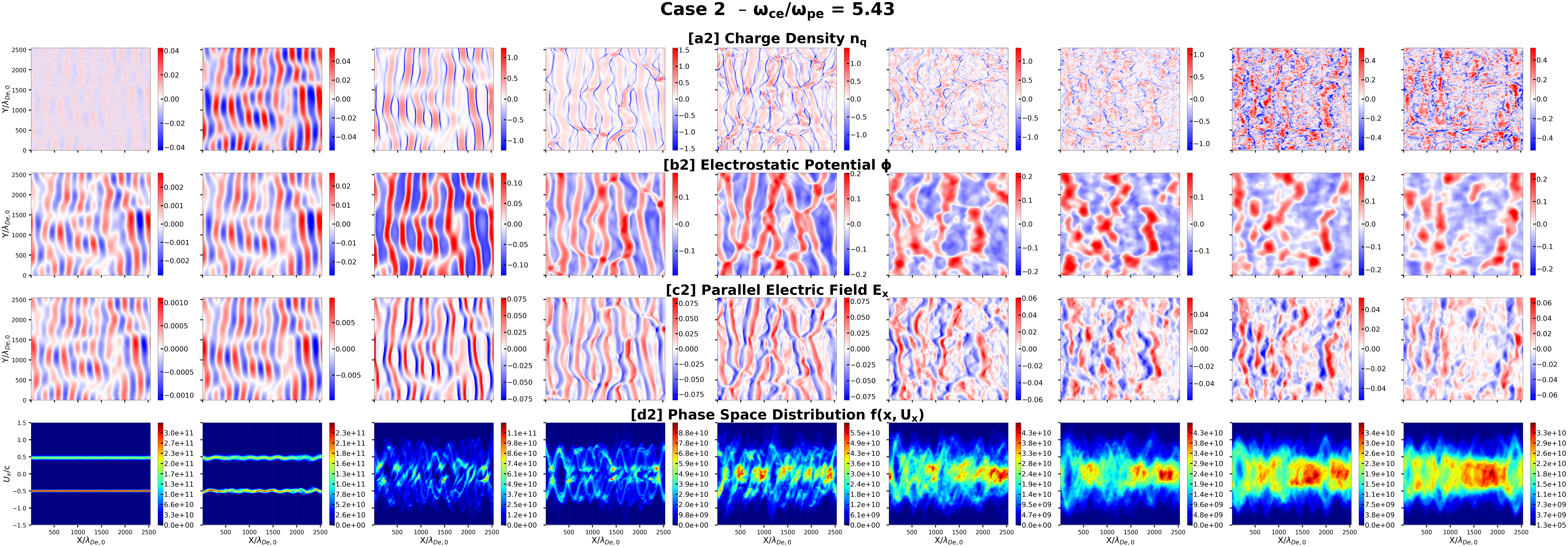}
\includegraphics[width=0.8\textwidth] {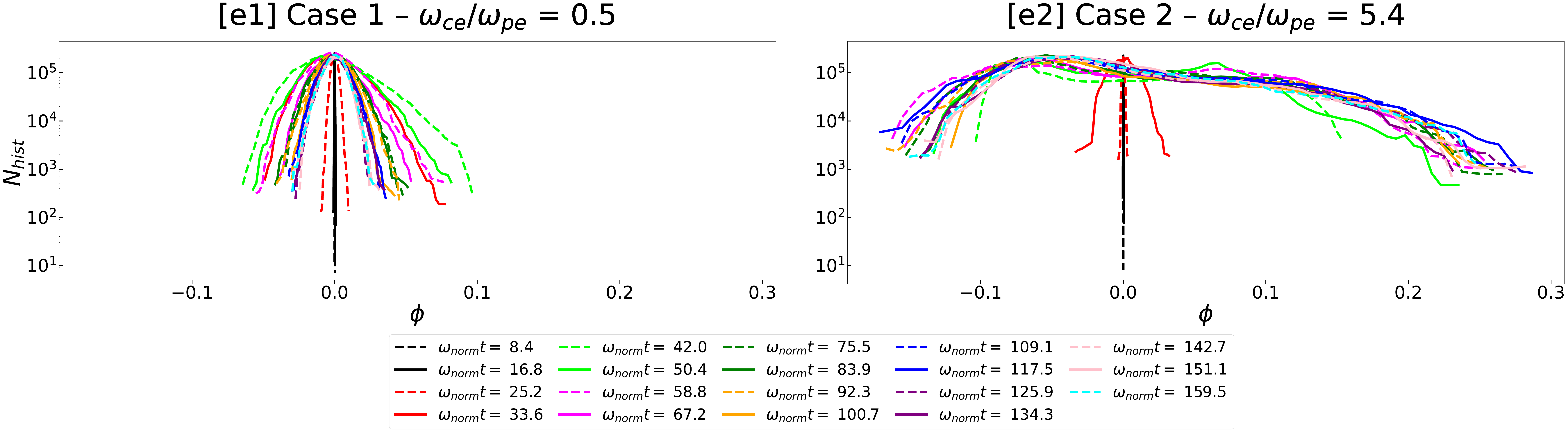}
\caption{Temporal evolution of the spatial distributions of parameters: the charge density $n_q$ (rows [$a1$] and [$a2$]), the potential $\phi$ (rows [$b1$] and [$b2$]), the parallel electric field $E_x$ (rows [$c1$] and [$c2$]), and the phase-space distribution $f(x,U_x)$ of all electrons (rows [$d1$] and [$d2$]) in two differently magnetized plasmas (Case 1 and Case 2 with $\omega_{ce}/\omega_{pe}=0.54$ and $5.4$, respectively). Each column corresponds to a time snapshot, as labeled at the top. Note that the position coordinates in each panel are normalized by $\lambda_{De}$, and each panel uses a different color scale.
Panels [$e1$] and [$e2$] display the histograms of the $\phi$ well values at different time snapshots for Case 1 and Case 2, respectively. 
}
\label{Fig1}
\end{center}
\end{figure*}

Figure~\ref{Fig1} shows the evolution of several key parameters in two differently magnetized electron bi-stream plasmas: the charge density $n_q$, the (electrostatic) potential $\phi$, the parallel electric field $E_{\parallel}$, and the electron phase-space distribution $f(x,U_x)$, with $\omega_{ce}/\omega_{pe} = 0.54$ (i.e., the weakly magnetized plasma, Case 1) and $5.4$ (i.e., the strongly magnetized plasma, Case 2), respectively. In both cases, the system starts from a bi-stream electron distribution and undergoes linear instability, nonlinear trapping, as well as the formation and evolution of phase-space vortices.

In rows [$d1$] and [$d2$] for $f(x,U_x)$, two separated electron beams are visible at early time $\omega_{norm}t \lesssim 33.6$. According to the definition of the EH given above, no EH has yet formed at this stage, although the positive $\phi$ humps are already significantly enhanced (see rows [b1] and [b2], or panels [e1] and [e2]). It should also be noted that within the entire simulation domain, both positive and negative $\phi$ structures appear alternately due to the global electron number conservation. Since both the positive and negative $\phi$ structures can trap electrons, they are hereinafter referred to as $\phi$ wells. 

As the instability develops (i.e., during $\omega_{norm}t \approx 42$--$58.8$), the beams broaden in velocity space and eventually develop well-defined low-density closed contours, i.e., trapped-electron phase-space vortices, around $\omega_{norm}t \approx 42$ and $\approx 58.8$ in the weakly and strongly magnetized plasmas, respectively. Simultaneously, localized coherent positive $\phi$ humps (rows [$b1$] and [$b2$]) reach their maxima, accompanied by pronounced electron-density depletion and positive $n_q$ peaks (rows [$a1$] and [$a2$]), as well as by strong bipolar $E_{\parallel}$ pulses (rows [$c1$] and [$c2$]). These four features together constitute the characteristic signature of an electron hole (EH): an electron density deficit, a localized positive potential, a bipolar parallel electric field, and a trapped-electron phase-space vortex. 
Hence, the well-developed EHs are manifested as mature positive $\phi$ humps.
The spatial morphology of these mature positive $\phi$ humps (i.e., EHs) depends strongly on the plasma magnetization. In the weakly magnetized case, the humps are patch-like and exhibit 2D structures, with their perpendicular extent (along the $y$-axis) generally larger than the parallel one. In the strongly magnetized plasma, they instead appear as elongated stripes aligned perpendicular to $\vec{B}_{0}$, i.e., 1D structures. Moreover, both the length scale along each spatial axis and the amplitude of these positive $\phi$ humps (panel [$e2$]) are larger in the strongly magnetized plasma, and the corresponding EH vortices extend further (exhibiting bifurcation structures) in the parallel-velocity ($U_{x}$) space as well.

At later times ($\omega_{norm}t \gtrsim 75.5$), the two plasmas exhibit distinctly different evolutionary behaviors. 
In the weakly magnetized plasma, the fine structures in the $n_q$ (row [$a1$]), positive $\phi$ humps (row [$b1$]), and $E_{\parallel}$ (row [$c1$]) become irregular, and the amplitudes of the $\phi$ wells decrease with time (panel [$e1$]). In the phase space (row [$d1$]), the EH vortices are gradually smeared out and eventually disappear, leaving a broadened and diffuse distribution along $U_x$. 
In the strongly magnetized plasma, by contrast, the positive $\phi$ humps eventually fragment into 2D patches (row [$b2$]). Although these patches resemble the well-developed $\phi$ humps in the weakly magnetized plasma at $\omega_{norm}t \sim 42$, they possess larger spatial scales. The amplitudes of the $\phi$ wells, however, remain nearly unchanged (panel [$e2$]). 
Larger amplitudes of the  $\phi$ wells imply that more energy is stored within these structures, leading simultaneously to their larger spatial scales and longer dissipation times (i.e., longer lifetimes).
The $n_q$ (row [$a2$]) and $E_{\parallel}$ (row [$c2$]) are likewise fragmented, with a slight decrease in intensity over time. In the phase space (row [$d2$]), the EH vortices persist and retain well-defined closed contours. The total number of EHs, however, decreases over time, possibly due to merging between neighboring vortices. 
Overall, the EHs can survive longer under strong magnetization, consistent with previous related PIC investigations~\citep[e.g.,][]{Singh_etal_2001JGR...10621183S, Wu_etal_2013PlST...15...17W, Hutchinson_2017PhPl...24e5601H}.

\subsection{wave excitation}

\begin{figure*}
\begin{center}
\includegraphics[width=0.8\textwidth]{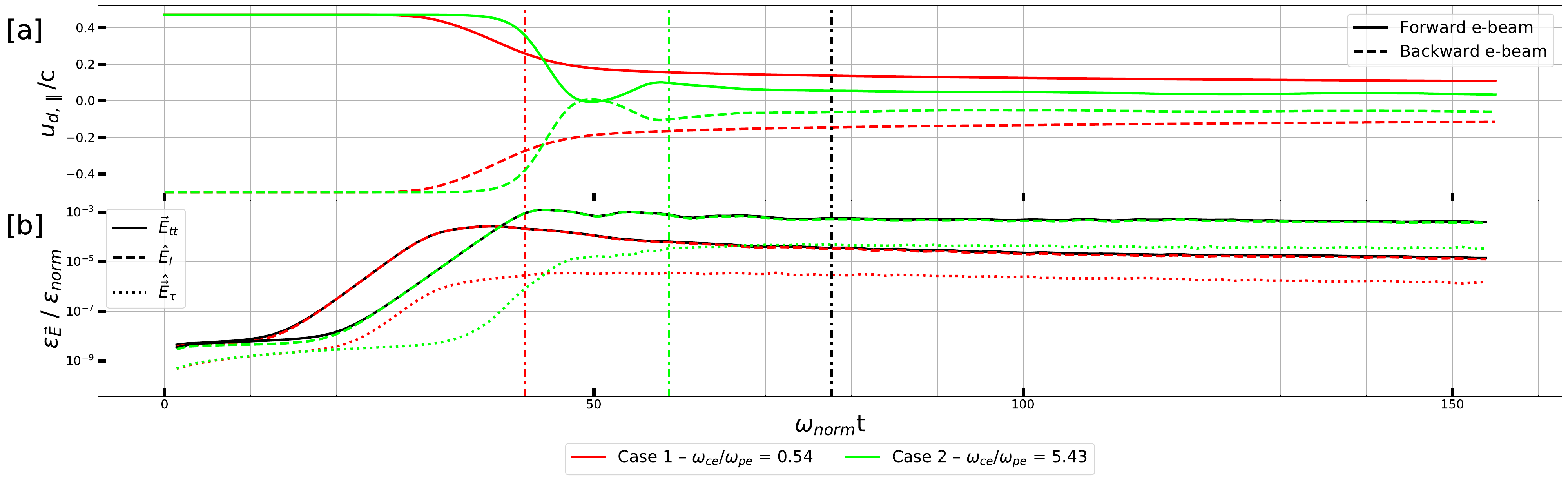}
\includegraphics[width=1.0\textwidth]{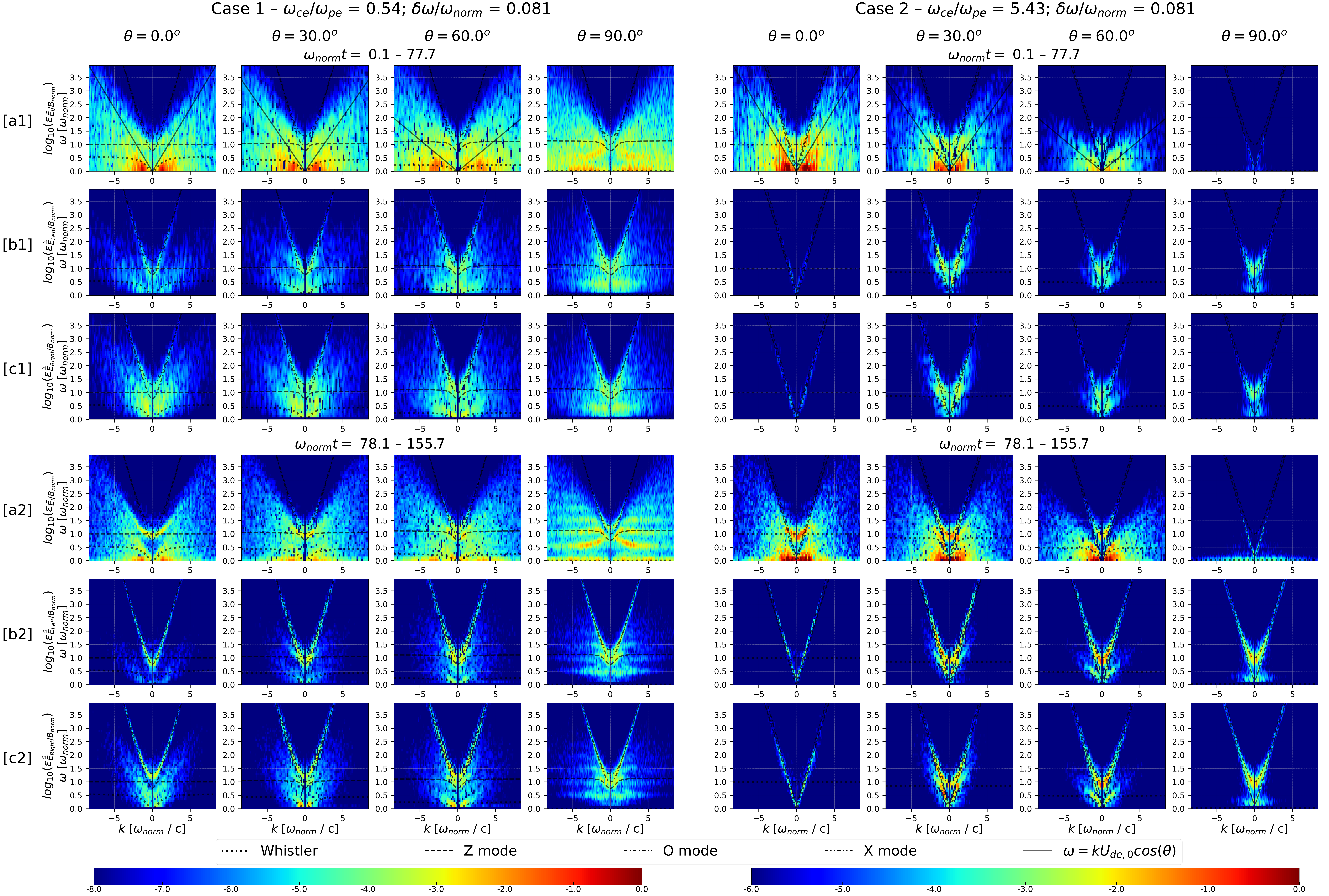}
\caption{Panel [$a$] presents the temporal evolution of the drift velocity of the electron bi-streams. Panel [$b$] shows the temporal evolution of the total ($\vec{E}_{tt}$, solid line), longitudinal ($\hat{E}_{l}$, dashed line), and transverse ($\hat{\vec{E}}_{\tau}$, dotted line) electric energy of waves in the whole simulation domain. In both panels [$a$] and [$b$], the two leftmost vertical dashed-dotted lines indicate the time snapshot at which the EHs are well developed in the two different plasmas, respectively; the third vertical dashed-dotted line is the boundary between the two time windows used for the dispersion relation power spectrum. Different colors in each panel are used to distinguish the different cases. 
The bottom six rows present the wavevector-frequency ($\vec{k}$--$\omega$ or dispersion relation) power spectrum of the longitudinal electric fields ($\tilde{E}_{l}$ in rows [$a1$], [$a2$]), as well as the left-hand ($\tilde{\vec{E}}_{Left}$ in rows [$b1$], [$b2$])  and right-hand ($\tilde{\vec{E}}_{Right}$ in rows [$c1$], [$c2$]) polarized transverse electric fields of waves in the whole simulation domain. The 1-series (rows [$a1$], [$b1$] and [$c1$]) and the 2-series (rows [$a2$], [$b2$] and [$c2$]) correspond to the time period $\omega_{norm} t = 0.1$--$77.7$  and $78.1$--$155.7$, respectively. Meanwhile, the left and right four columns display the $\vec{k}$--$\omega$ power spectra for Case 1 and 2, respectively, along four different wave propagation angles $\theta$.  
In each panel, from bottom to top, the overplotted dotted, dashed, dashed-dotted, and dashed-dotted-dotted lines represent the theoretical dispersion relations of the Whistler, Z, O, and X modes, respectively, in the magnetized cold-plasma limit. The solid lines in row [$a1$] correspond to the dispersion relation of the beam mode. Additionally, in eqach panel, $|k|^{2}=k_{\parallel}^{2}+k_{\perp}^{2}$, and the sign of $k$ is the same as that of its parallel component $k_{\parallel}$. The frequency resolution (i.e., $\delta \omega$) in these spectra is $0.081 \omega_{norm}$. 
}
\label{Fig2}
\end{center}
\end{figure*}

Figure~\ref{Fig2} provides a overview of the temporal evolution of the electron stream's drift velocity (panel [$a$]) and different electric energies (panel [$b$]), as well as the dispersion relation power spectra of excited waves in the system (bottom six rows). In panel [$b$], the total ($\vec{E}_{tt} = [E_{x}(x, y, t), E_{y}(x, y, t), E_{z}(x, y, t)]$), longitudinal ($\hat{E}_{l} = \hat{\vec{E}}(\vec{k}, t) \cdot \vec{k}/|k| $), and transverse ($\hat{\vec{E}}_{\tau} = \hat{\vec{E}}(\vec{k}, t) \times \vec{k}/|k|$) electric energies are denoted by solid, dashed, and dotted lines, respectively. $|\vec{E}_{tt}|^{2} = E_{x}(x, y, t)^{2} + E_{y}(x, y, t)^{2} + E_{z}(x, y, t)^{2} = \hat{E}_{l}^{2} + |\hat{\vec{E}}_{\tau}|^2$.  
Note that in this paper, longitudinal and transverse electric fields are referred to as electrostatic and electromagnetic waves, respectively, although some obliquely propagating electromagnetic waves (e.g., the whistler, Z-, O-, and X-mode waves~\citealp{Willes&Cairns_2000PhPl....7.3167W}) also possess a longitudinal electric field component.

As shown in panels [$a$] and [$b$] of Fig.~\ref{Fig2}, the energies of both the electrostatic and electromagnetic waves increase with decreasing drift velocity of the electron streams. This is consistent with the theoretical prediction of the bi-stream instability: the reduction of the drift velocity prolongs the wave--particle interaction, transferring their energy to waves. Moreover, the decrease in the drift velocity, as well as the energy gains of both the electrostaic and electromagnetic waves, is delayed but larger in the strongly magnetized plasma than in the weakly magnetized one. Furthermore, panel [$b$] of Fig.~\ref{Fig2} reveals that the electrostatic waves grow more rapidly and saturate earlier with an energy roughly an order of magnitude higher than that of the electromagnetic waves in both plasmas. This indicates that the reduced beam energy is preferentially converted into the electrostatic waves, regardless of the background magnetization.

Comparing the time evolution of the EHs (Fig.~\ref{Fig1}) with the energy evolution of waves (panel [$b$] of Fig.~\ref{Fig2}) reveals a distinct temporal sequence for both plasmas: the electrostatic waves saturate before the moment $t_{EHm}$ when the $\phi$ wells are mature and the EHs become well developed, specifically at $\omega_{norm} t_{EHm} \approx 42$ in the weakly magnetized plasma (red dashed-dotted line in panel [$b$] of Fig.~\ref{Fig2}) and $\omega_{norm} t_{EHm} \approx 58.8$ in the strongly magnetized one (lime dashed-dotted line in panel [$b$] of Fig.~\ref{Fig2}), since the EHs are the nonlinear products of the electrostatic waves. 
For the electromagnetic waves, their excitation starts earlier than $t_{EHm}$, but their saturation lags slightly behind $t_{EHm}$. By contrast, the time at which the electromagnetic wave energy increases significantly corresponds well to that at which the amplitude of the positive $\phi$ humps are significantly enhanced at $\omega_{norm} t \approx 20$ in the weakly magnetized plasma and at $\omega_{norm} t \approx 30$ in the strongly magnetized one (panels [$e1$] and [$e2$] of Fig.~\ref{Fig1}). This indicates that the excitation of the electromagnetic waves is strongly related to the $\phi$ wells rather than to the EHs.

To determine the wave modes of the excited electrostatic and electromagnetic waves, the rows [$a1$] to [$c2$] of Fig.~\ref{Fig2} display the dispersion relation ($\vec{k}$--$\omega$) power spectra of waves propagating along different directions $\theta = \arctan(k_{\perp}/k_{\parallel})$, shown in different columns. Here $k_{\parallel} = \vec{k} \cdot \vec{B}_{0} / |B_{0}|$ and $k_{\perp} = |\vec{k} \times \vec{B}_{0} | / |B_{0}|$. These spectra are temporally averaged over the whole simulation domain for two periods: $\omega_{norm} t = 0.1$--$77.7$ (rows [$a1$], [$b1$], [$c1$])  and $\omega_{norm} t = 78.1$--$155.7$ (rows [$a2$], [$b2$], [$c2$]). The time period $\omega_{norm} t = 0.1$--$77.7$ covers the linear-growth phases of all excited waves, whereas the plasmas reach a nonlinear quasi-equilibrium phase after $\omega_{norm} t = 78$. 
Rows [$a1$] and [$a2$] show the spectra of the longitudinal electric fields ($\tilde{E}_{l} = \tilde{\vec{E}} \cdot \vec{k} / |k|$) for the electrostatic waves, where $\tilde{\vec{E}}$ is the Fourier transforms of the real-space electric field $\vec{E}(x, y, t)$. Rows [$b1$]/[$b2$] and [$c1$]/[$c2$] display the left-handed ($\tilde{\vec{E}}_{Left}$) and right-handed ($\tilde{\vec{E}}_{Right}$) polarized transverse electric fields, respectively, with respect to $\vec{k}$, for the electromagnetic waves. Note that $\tilde{\vec{E}}_{\tau} =  \tilde{\vec{E}} \times \vec{k} / |k| = 0.5(\tilde{\vec{E}}_{Left} + \tilde{\vec{E}}_{Right})$ (see~\citealp{Zhou_etal_2020ApJ...891...92Z} for more details about the method to separate the left-handed and right-handed polarized electromagnetic waves).
The overplotted lines in each panel denote the theoretical dispersion relations of waves in the magnetized cold plasma limit~\citep[see, e.g.,][]{Melrose_1986islp.book.....M, Stix_1992wapl.book.....S, Willes&Cairns_2000PhPl....7.3167W} as a reference. By comparing the simulated spectra with the theoretical dispersion relation lines, it is evident that the excited waves can be classified as follows.

\subsubsection{Electrostatic waves}
\label{ES}

Initially, in both plasmas, due to the electron beam instability driven by the bi-stream electrons (see Section 3.2.2 in~\citealp{Gary_1993tspm.book.....G}), the first and most efficiently excited waves are the electrostatic beam-mode waves, which are excited via a linear process, with energies primarily concentrated in the low-frequency ($\omega < 0.5 \omega_{pe}$) region.
These excited electrostatic beam-mode waves exhibit a broad propagation-angle range in the weakly magnetized plasma (see the left four panels of row [$a1$]of Fig.~\ref{Fig2}), while they are quasi-parallel propagating in the strongly magnetized plasma (see the right four panels of row [$a1$] of Fig.~\ref{Fig2}). The weak or absent excitation of electrostatic waves along oblique directions in the strongly magnetized plasma is probably due to the stabilization of the transverse instability by the strong ambient magnetic field~\citealp{Wu_etal_2013PlST...15...17W}.
At the initial stage of the system evolution (i.e., $\omega_{norm} t < 33.6$, see rows [$b1$] and [$b2$] of Fig.~\ref{Fig1}), the wave fronts of these electrostatic beam-mode waves correspond to the edges of the alternating positive and negative $\phi$ wells. 
Reflecting the propagation anisotropies of the beam-mode waves, during this initial stage, the $\phi$ wells in the weakly magnetized plasma show meandering channels, while those in the strongly magnetized plasma present as elongated stripes aligned perpendicular to $\vec{B}_{0}$.

During the early stage of the electron bi-stream system evolution (i.e., $\omega_{norm} t  = 0.1$--$77.7$), besides the electrostatic beam-mode waves, there could also be energized electrostatic whistler waves, as well as broadly propagating ultra-low-frequency electrostatic waves with frequencies near or below the frequency resolution of these $\vec{k}$--$\omega$ spectra, $\delta \omega/\omega_{norm}=0.081$, in both plasmas.
Based on previous investigations of EHs~\citep{Omura_etal_1996JGR...101.2685O, Singh_etal_2001JGR...10621183S}, the electrostatic whistler and ultra-low-frequency waves are the decay products of the EHs. Specifically, the ultra-low-frequency electrostatic waves could be ion acoustic~\citep{Omura_etal_1996JGR...101.2685O} and/or lower hybrid~\citep{Singh_etal_2001JGR...10621183S} waves. In the following context, we will concentrate on the waves with $\omega > \delta \omega$ only.
Additionally, perpendicularly propagating electrostatic Bernstein harmonics are also present in the weakly magnetized plasma, first reported in this study electron stream plasmas.

As time evolves into the nonlinear quasi-equilibrium stage (i.e., $\omega_{norm} t = 78.1$--$155.7$, see row [$a2$] of Fig.~\ref{Fig2}), accompanied by the energy decrease of the electrostatic waves with $\omega < 0.5 \omega_{pe}$ in the weakly magnetized plasma and the quasi-parallel propagating beam-mode waves in the strongly magnetized plasma, the (spectral) energies of the quasi-parallel propagating Langmuir waves and obliquely propagating electrostatic Z-mode waves increase in both plasmas. 
Additionally, obliquely propagating electrostatic whistler waves in the strongly magnetized plasma and the perpendicularly propagating electrostatic Bernstein harmonics in the weakly magnetized plasma show energy gains as well. 
As pointed out by~\citealp{Singh_etal_2001GeoRL..28.1371S}, the electrostatic energy transfer from the quasi-parallel direction to the oblique directions in the strongly magnetized plasma is related to the real-space structure transition (from the initial 1D to 2D structures) of the $\phi$ wells and EHs (see row [$b2$] in Fig.~\ref{Fig1}). 
It should be noted that because the spectral energy is obtained by integrating (or averaging) the wave energy over a specific time period, a higher spectral energy in the later (time) window does not imply that the wave is still being excited. The wave could have already been fully excited (growing in energy) in the previous window and simply remained at a constant energy level in the subsequent window.

\subsubsection{Electromagnetic waves} 

First note that, with respect to the wavevector $\vec{k}$, the polarization of an electromagnetic wave changes with $\theta$~\citep{Melrose_1986islp.book.....M}. For instance, (1) if $\theta = 0^{\circ}$, the O-mode (whistler and X-mode) waves are fully left-hand (right-hand) circularly polarized, and the Z-mode waves are fully left-hand (right-hand) circularly polarized when their frequencies $\omega < (>)~\omega_{pe}$; (2) if $\theta = 90^{\circ}$, the frequencies of the whistler waves are close to 0, while the Z-, O-, and X-mode waves are linearly polarized; (3) these electromagnetic waves are elliptically polarized when they propagate obliquely (i.e., $\theta$ is neither $0^{\circ}$ nor $90^{\circ}$), which indicates that these electromagnetic waves contain a longitudinal electrc field component. Therefore, to more easily distinguish the modes of the excited electromagnetic waves, the left- and right-hand polarized components are displayed separately in the dispersion spectra of Fig.~\ref{Fig2}. 

Consistent with panel [$b$] of Fig.~\ref{Fig2}, the energies of the excited electromagnetic waves are lower than those of the electrostatic waves in both plasmas, and the electromagnetic excitation is stronger in the strongly magnetized plasma. 
Unlike the excited electrostatic waves, most of the energies of the electromagnetic waves are located in directions non-parallel to $\vec{B}_{0}$ in both plasmas. In addition, these excited electromagnetic waves generally have a wider propagation range in the weakly magnetized plasma than in the strongly magnetized plasma.

Along propagation directions non-perpendicular to $\vec{B}_{0}$ (see the columns with $\theta \neq 90^{\circ}$ in rows [$b1$], [$c1$], [$b2$], and [$c2$] of Fig.~\ref{Fig2}), the excited electromagnetic waves are mainly right-hand polarized low-frequency whistler waves, as well as Z, O, and/or X modes near the plasma frequency $\omega_{pe}$ (the X mode is not efficiently excited in the strongly magnetized plasma, where its frequencies are all much larger than $\omega_{pe}$). The mixture of the Z, O, and/or X modes around $\omega_{pe}$ implies that the electromagnetic emission around $\omega_{pe}$ might be weakly polarized.

Besides the excitation strength of the electromagnetic waves, another remarkable difference between the two differently magnetized plasmas is that, in the direction perpendicular to $\vec{B_{0}}$ (i.e., $\theta = 90^{\circ}$), energy-enhanced electromagnetic harmonic electron cyclotron waves appear in the weakly magnetized plasma, corresponding to the excitation of the electrostatic Bernstein harmonics there; while in the strongly magnetized plasma, the excited electromagnetic modes are still mainly the O mode near $\omega_{pe}$ (similar to those propagating with $\theta \neq 90^{\circ}$).

Regarding the temporal evolution of the (spectral) energy, most excited electromagnetic waves show higher energy in the nonlinear quasi-equilibrium stage (i.e., $\omega_{norm} t = 78.1$--$155.7$) than in the earlier stage, except for those with frequencies $\delta \omega < \omega < \omega_{pe}$ in the weakly magnetized plasma. As mentioned in Sec.~\ref{ES}, the energies of the excited electrostatic waves with $\omega < 0.5 \omega_{pe}$ also decreas during the nonlinear quasi-equilibrium stage in the weakly magnetized plasma.
These spectral behaviors indicate the energy loss of these low-frequency waves. 
In the following, we focus on the excitation mechanism of these exited electrostatic and electromagnetic waves.

\subsection{Excitation mechanism of the electromagnetic waves} 
\label{mechanism}

From the perspective of classical electrodynamics (Maxwell's equations), the excitation of any wave mode in a plasma fundamentally originates from perturbations that break the quasi-neutral equilibrium of the plasma. 
Specifically, charge separation (or density perturbations) acts as the source for electrostatic waves. In contrast, current perturbations can drive electrostatic and/or electromagnetic waves; these perturbations include linear conduction currents (first-order responses to the perturbed fields, e.g., those induced by non-Maxwellian particle distributions) and nonlinear beating currents (second-order responses driven by wave-wave coupling, e.g., coupling currents arising from wave-wave interactions).
Ultimately, from a kinetic perspective, all kinds of current perturbations originate from the relative drift motions of charged particles.

Due to the presence of (both positive and negative) $\phi$ wells in the electron bi-stream plasmas (see rows [$b1$] and [$b2$] of Fig.~\ref{Fig1}), electrons can undergo bounce oscillations within the $\phi$ wells~\citealp{Bernstein_1957PhRv..108..546B} and perpendicular $\vec{E}_{\perp} \times \vec{B}_{0}$ drifts~\citealp{Umeda_etal_2004JGRA..109.2207U, Wu_etal_2013PlST...15...17W}. Since the total electric energy is dominated by the electrostatic component in both plasmas (see panel [$b$] of Figure~\ref{Fig2}), the perpendicular electric field is approximated by $\vec{E}_{\perp} \approx E_{y} \hat{y}\approx - \frac{\partial \phi}{\partial y} \hat{y}$. 
These motions, in turn, lead to localized currents (together with other localized components, such as inductive currents induced by the time-varying local charge distribution, and diamagnetic currents driven by density gradients at the $\phi$ well boundaries) that serve as sources to excite electrostatic and electromagnetic waves.

\begin{figure*}
\begin{center}
\includegraphics[width=1.0\textwidth]{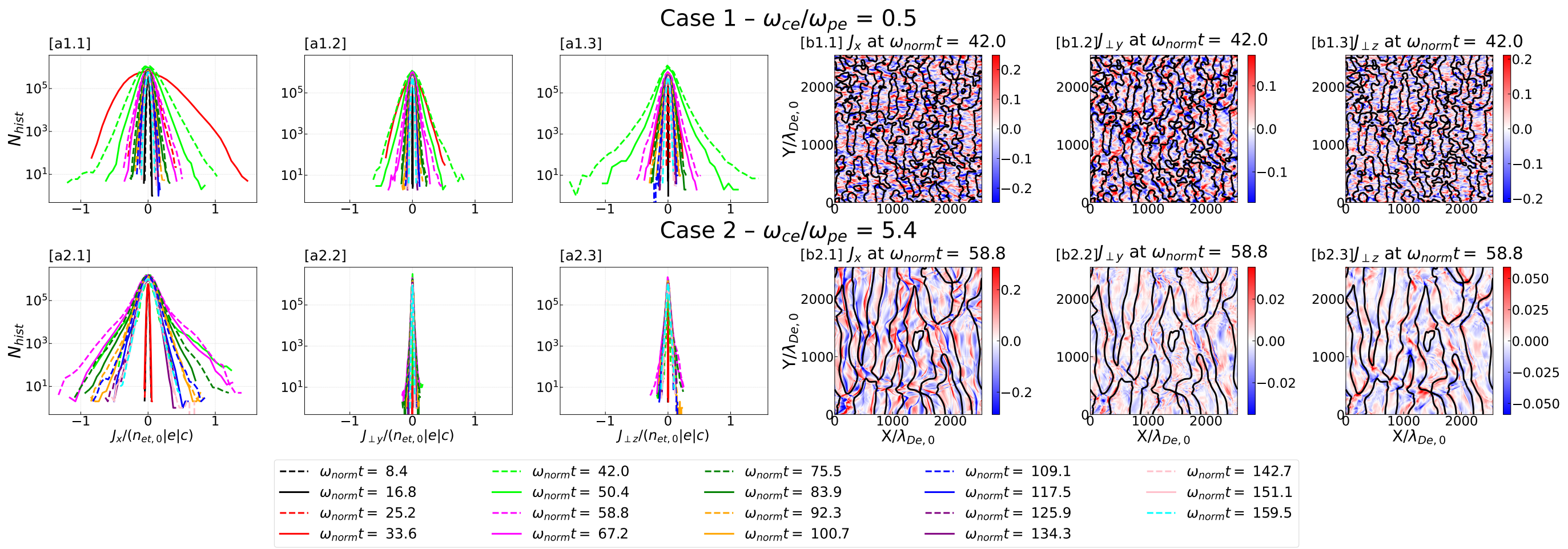}
\caption{Temporal evolution of the histograms of the current density components and their spatial distributions at the well-developed stage of the EHs in the weakly and strongly magnetized plasmas. [$a1.1$]--[$a1.3$]: Temporal evolution of the current density component  $(J_{x}$, $J_{\perp y}$, and $J_{\perp z})$ histograms in the weakly magnetized plasma (Case 1), where $n_{et,0}$ is the initial number density of all electrons, and $e$ is the elementary charge (a positive constant). [$a2.1$]--[$a2.3$]: Same as [$a1.1$]--[$a1.3$], but for the strongly magnetized plasma (Case 2). [$b1.1$]--[$b1.3$]: Spatial distributions of the corresponding current density components in the weakly magnetized plasma at $\omega_{norm} t_{EHm} = 42.0$. The black contour lines represent the zero-potential boundaries ($\phi = 0$). [$b2.1$]--[$b2.3$]: Same as [$b1.1$]--[$b1.3$], but for the strongly magnetized plasma at $\omega_{norm} t_{EHm} = 58.8$.
}
\label{Fig3}
\end{center}
\end{figure*}

Figure~\ref{Fig3} presents the histogram temporal evolution and spatial distribution of the current density over the entire simulation domain for both the weakly (top two rows) and strongly (bottom two rows) magnetized plasmas. Panels [$a1.1$]--[$a1.3$] and [$a2.1$]--[$a2.3$] display the temporal evolution of the histograms for the three current density components $\vec{J} = (J_x, J_{\perp y}, J_{\perp z})(x, y, t)$ over the entire spatial domain, corresponding to [$a1.1$]/[$a2.1$], [$a1.2$]/[$a2.2$], and [$a1.3$]/[$a2.3$], respectively. Panels [$b1.1$]--[$b1.3$] and [$b2.1$]--[$b2.3$] show the corresponding spatial distributions of these current density components at the time snapshot when the EHs are well developed (i.e., at $t_{EHm}$) in the weakly and strongly magnetized plasmas, respectively. In these spatial distribution panels, the black contour lines indicate the regions where $\phi = 0$.

Panels [$a1.1$]--[$a1.3$] and [$a2.1$]--[$a2.3$] of Fig.~\ref{Fig3} reveal that all current density components reach their maxima around $t_{EHm}$. The peak parallel current densities ($J_x$) are comparable in both plasmas. A significant difference lies in the transverse current densities ($J_{\perp y}$, $J_{\perp z}$): they are strongly enhanced in the weakly magnetized plasma (even comparable to $J_x$), but remain nearly zero in the strongly magnetized case. This difference is probably due to the strong magnetic field suppressing the perpendicular motion of electrons.
Furthermore, coincidence with the amplitude evolution of the $\phi$ wells (see panels [$e1$] and [$e2$] of Fig.~\ref{Fig1}), all current density components decay to zero along with the dissipation of the $\phi$ wells in the weakly magnetized plasma, while $J_x$ in the strongly magnetized plasma is persistently maintained at relative high values due to the stable $\phi$ wells there. 
Most importantly, the spatial distributions of all current density components highly coincide with the structures of the $\phi$ wells (indicated by the black contours) at $t_{EHm}$ (see panels [$b1.1$]--[$b1.3$] and [$b2.1$]--[$b2.3$] of Fig.~\ref{Fig3}).
All these features collectively demonstrate that these localized current sources originate directly from the dynamics of the electrons trapped by the $\phi$ wells. Unlike wave-induced currents that usually exhibit prominent spatial coherence, periodicity, ordered patterns, and remain sustained throughout the nonlinear quasi-equilibrium stage, these trapped-electron currents are instead localized, spatially confined, and evolve continuously together with the $\phi$ wells over the entire simulation time period.

\begin{figure*}
\begin{center}
\includegraphics[width=1.0\textwidth]{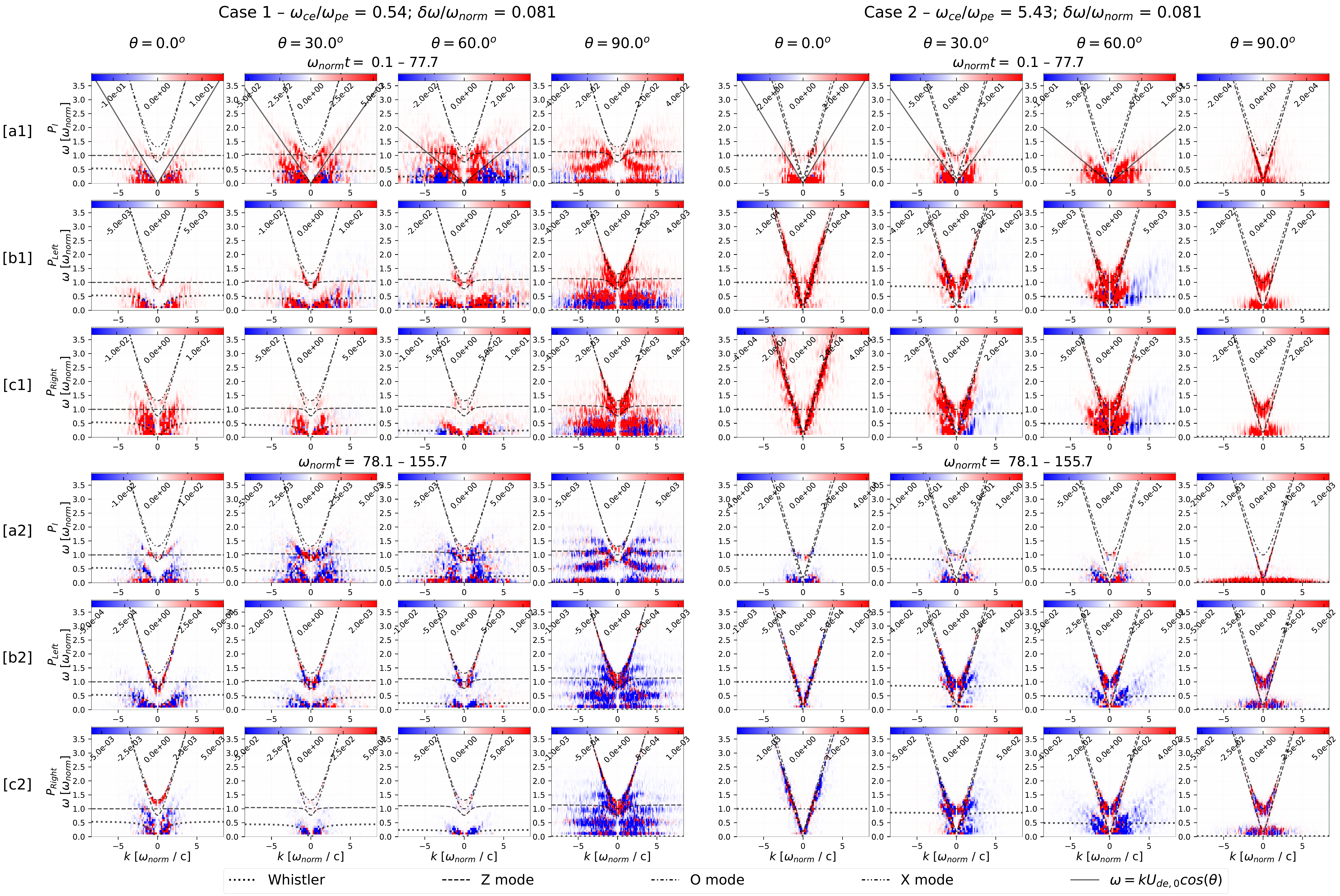}
\caption{Similar to the dispersion relation power spectra in Fig.~\ref{Fig2}, spectra of the energy transfer rate,  $P = -\mathrm{Re}[\tilde{\vec{J}}^{*} \cdot \tilde{\vec{E}}]$, are shown for three polarization components: longitudinal ($P_{l} = -\mathrm{Re}[\tilde{\vec{J}}^{*}_{l} \cdot \tilde{\vec{E}}_{l}]$ in rows [$a1$] and [$a2$]), left-handed ($P_{Left} = -\mathrm{Re}[\tilde{\vec{J}}^{*}_{Left} \cdot \tilde{\vec{E}}_{Left}]$ in rows [$b1$] and [$b2$]), and right-handed ($P_{Right} = -\mathrm{Re}[\tilde{\vec{J}}^{*}_{Right} \cdot \tilde{\vec{E}}_{Right}]$ in rows [$c1$] and [$c2$]) over different time periods in differently magnetized plasmas. 
}
\label{Fig4}
\end{center}
\end{figure*}

Based on the Poynting theorem, an effective wave excitation requires the current sources to be spatiotemporally matched with the dispersion relation and polarization of the target wave modes. The energy transfer rate between currents and waves is given by $P= -\mathrm{Re}[\tilde{\vec{J}}^{*} \cdot \tilde{\vec{E}}]$ in the wavevector-frequency ($\vec{k}, \omega$) space, where $\tilde{\vec{J}}$, as with $\tilde{\vec{E}}$, is the Fourier transform of the current density $\vec{J}$ in real space; $\tilde{\vec{J}}^{*}$ is the complex conjugate of $\tilde{\vec{J}}$. A significantly positive value of $P$ indicates that the current efficiently pumps energy into waves. A negative $P$, however, implies energy loss from waves. The magnitude of $P$ (i.e., $|P|$) reflects the strength of the excitation or damping.

Figure~\ref{Fig4} displays the spectra for the three polarization components of $P$, i.e., the longitudinal $P_{l} = -\mathrm{Re}[\tilde{\vec{J}}^{*}_{l} \cdot \tilde{\vec{E}}_{l}]$, left-handed $P_{Left} = -\mathrm{Re}[\tilde{\vec{J}}^{*}_{Left} \cdot \tilde{\vec{E}}_{Left}]$, and right-handed $P_{Right} = -\mathrm{Re}[\tilde{\vec{J}}^{*}_{Right} \cdot \tilde{\vec{E}}_{Right}]$ components, over the entire simulation domain for different time periods $\omega_{norm} t = 0.1$--$77.7$ and $78.1$--$155.7$ in different plasmas.
In both plasmas, the early stage (i.e., $\omega_{norm} t = 0.1$--$77.7$ in rows [$a1$], [$b1$], and [$c1$] of Fig.~\ref{Fig4}) features a uniformly positive $P$ along the dispersion curves of the excited waves (see rows [$a1$] to [$c2$] in Fig.~\ref{Fig2}), while sign reversals are prominent in the nonlinear quasi-equilibrium stage (i.e., $\omega_{norm} t = 78.1$--$155.7$ in rows [$a2$], [$b2$], and [$c2$] of Fig.~\ref{Fig4}). The regions of mixed positive and negative $P$ indicate a dynamic energy balance: the waves are no longer net amplified, instead, energy is either nonlinearly transferred between different modes (e.g., from the mother wave to the daughter waves) or oscillating between waves and electrons. 
Besides these mixed regions, there are still some regions with $P > 0$ during the nonlinear quasi-equilibrium stage, e.g., the parallel propagating right-hand polarized X mode in the weakly magnetized plasma (see the leftmost panel in row [$c2$] of Fig.~\ref{Fig4}). Considering the weak current density components after $\omega_{norm} t = 78$ (see panels [$a1.1$]--[$a1.3$] in Fig.~\ref{Fig3}), the excitation of this parallel-propagating X mode during the nonlinear quasi-equilibrium stage is likely attributed to wave-wave interactions (e.g., two counter-parallel-propagating whistler waves with $P< 0$) .

\begin{figure*}
\begin{center}
\includegraphics[width=1.0\textwidth]{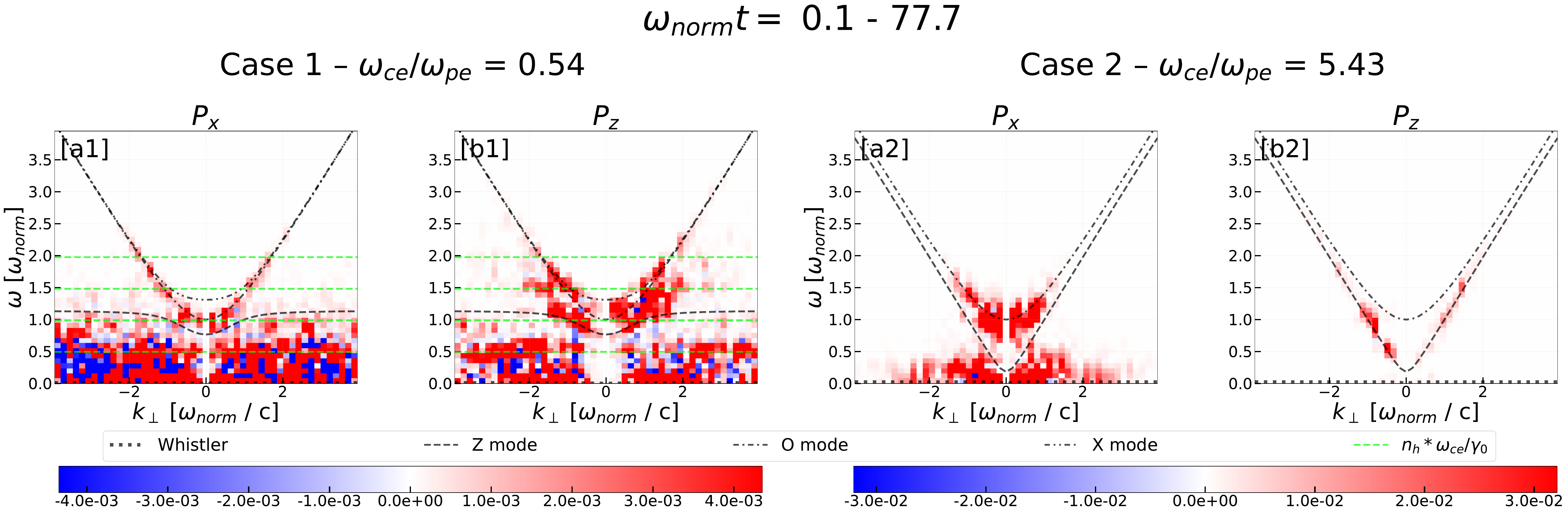}
\caption{Similar to Fig.~\ref{Fig4}, spectra of $P_{x} = -\mathrm{Re}[\tilde{J}^{*}_{x} \cdot \tilde{E}_{x}]$ (panels [$a1$] and [$a2$]) and $P_{z} = -\mathrm{Re}[\tilde{J}^{*}_{z} \cdot \tilde{E}_{z}]$ (panels [$b1$] and [$b2$]) along the perpendicular direction during $\omega_{norm} t = 0.1$--$77.7$ in differently magnetized plasmas. The lime dashed horizontal lines correspond to different harmonics of the electron gyrofrequency, i.e., $n_{h} \omega_{ce} / \gamma_{0}$.}
\label{Fig5}
\end{center}
\end{figure*}

For the excitation of the perpendicularly propagating electrostatic Bernstein and electromagnetic electron cyclotron harmonics in the weakly magnetized plasma, based on the Poynting theorem, one can deduce that $P_{y} = -\mathrm{Re}[\tilde{J}^{*}_{y} \cdot \tilde{E}_{y}]$ contributes to the excitation of the electrostatic Bernstein harmonics, while $P_{x} = -\mathrm{Re}[\tilde{J}^{*}_{x} \cdot \tilde{E}_{x}]$ and $P_{z} = -\mathrm{Re}[\tilde{J}^{*}_{z} \cdot \tilde{E}_{z}]$ contribute to the electromagnetic harmonic electron cyclotron waves. Note that there $P_{x} + P_{z} = P_{Left} + P_{Right}$ for perpendicularly propagating electromagnetic waves.
Figure~\ref{Fig5} shows separately the contributions of $P_{x}$ (panels [$a1$] and [$a2$]) and $P_{z}$ (panels [$b1$] and [$b2$]) to the excitation of these perpendicularly propagating electromagnetic waves in the different plasmas. 
For the waves with higher frequencies, as shown in Figure~\ref{Fig5}, $P_{x}$ contributes to the excitation of the O-mode waves, while $P_{z}$ is mainly responsible for the excitation of the Z- and X-mode waves, as well as the electromagnetic harmonic electron cyclotron waves. 
Therefore, the excitations of both the electrostatic Bernstein harmonics and electromagnetic electron cyclotron harmonics in the weakly magnetized plasma originate from the perpendicular currents. 
This excitation mechanism driven by perpendicular currents (termed the electron cyclotron drift instability) has indeed been confirmed by previous theoretical studies~\citep[e.g.,][]{Brinca_etal_2003JGRA..108.1038B, Muschietti_etal_2013JGRA..118.2267M, Janhunen_etal_2018PhPl...25h2308J}. 

\begin{figure*}
\begin{center}
\includegraphics[width=1.0\textwidth]{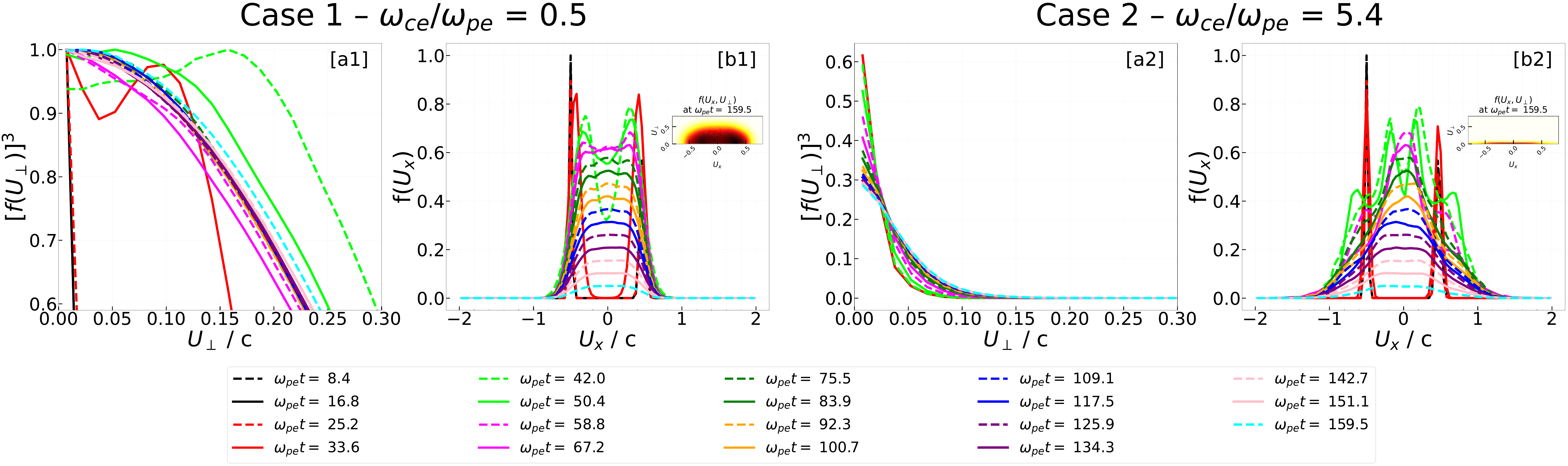}
\caption{Temporal evolution of electron velocity distributions of all electrons in the weakly and strongly magnetized plasmas. [$a1$] and [$a2$]: Perpendicular velocity distributions, presented as $[f(U_{\perp})]^{3}$. [$b1$] and [$b2$]: Corresponding parallel velocity distributions $f(U_x)$. The insets in panels [$b1$] and [$b2$] present the two-dimensional velocity distributions $f(U_x, U_{\perp})$ near the end of the respective simulations, i.e., at $\omega_{norm} t = 159.5$.
}
\label{Fig6}
\end{center}
\end{figure*}

In the kinetic theory of plasmas, current is determined by the velocity distributions of particles, i.e., $ \vec{J}(\vec{r}, t) = \sum_{s} q_{s} \int \vec{U} f_{s}(\vec{r}, \vec{U}, t) \, d^{3}\vec{U}$, where $s$ denotes the particle species, $q_{s}$ is the charge of species $s$, and $f_{s}(\vec{r}, \vec{U}, t)$ is the velocity distribution of species $s$ at position $\vec{r}$ and time $t$. Here, $\int f_{s}(\vec{r}, \vec{U}, t) \, d^{3}\vec{U} = n_{s}(\vec{r}, t)$ and $\int n_{s}(\vec{r}, t) \, d^{3}\vec{r} = N_{s}$, where $N_{s}$ is the total particle number of species $s$, which is conserved in the present simulations.
Throughout all the simulations, the protons still mainly retain their initial isotropic Maxwellian distribution. The current analyzed above is, hence, related to the velocity distribution functions of the electrons.

Panels [$a1$] and [$a2$] of Figure~\ref{Fig6} illustrate the temporal evolution of the perpendicular velocity distributions of all electrons in the weakly and strongly magnetized plasmas, respectively, where the distributions are cubed (i.e., $[f(U_\perp)]^{3}$) to clearly reveal their gradients. Panels [$b1$] and [$b2$] present the corresponding parallel velocity distributions $f(U_x)$. Note that the perpendicular distributions are first normalized by their maxima in panel [$a1$] and by the total electron number in panel [$a2$], and then cubed, whereas the parallel distributions in panels [$b1$] and [$b2$] are first normalized by their maxima and then shifted downward in chronological order to avoid overlap and better present the details of the distributions.

The positive gradient in the perpendicular velocity distribution (i.e., $\partial f / \partial U_\perp > 0$) is explicitly highlighted by the cubed plots in panel [$a1$] of Fig.~\ref{Fig6}. Physically, this positive gradient is a direct consequence of the symmetry breaking in velocity space induced by the perpendicular drift of electrons (e.g., the $\vec{E}_{\perp} \times \vec{B}_{0}$ drift), which simultaneously generates a sufficiently large perpendicular current (as shown in panels [$a1.2$] and [$a1.3$] of Fig.~\ref{Fig3} for the weakly magnetized plasma). Therefore, the positive gradient in velocity space and the large perpendicular current in real space represent two sides of the same physical process, which serves as a typical free-energy source for the excitation of the electrostatic Bernstein harmonics~\citep{Benacek&Karlicky2019ApJ...881...21B} and the electromagnetic harmonic electron cyclotron waves~\citep{Zhou_etal_2020ApJ...891...92Z}.
In contrast, the perpendicular velocity distribution in the strongly magnetized plasma lacks such a positive gradient and merely exhibits a slightly thermalized perpendicular Maxwellian distribution (panel [$a2$] of Fig.~\ref{Fig6}), as the strong magnetic field suppresses the perpendicular drifts of electrons. Therefore, the strongly magnetized plasma lacks both the strong perpendicular currents (panels [$a2.2$] and [$a2.3$] in Fig.~\ref{Fig3}) and the efficient excitation of the electrostatic Bernstein and electromagnetic electron cyclotron harmonics (see the $\vec{k}$--$\omega$ power spectra in Fig.~\ref{Fig2}).

In the strongly magnetized plasma, the parallel velocity distribution $f(U_x)$ (panel [$b2$] of Fig.~\ref{Fig6}), however, exhibits a distinct four-peak structure around $\omega_{norm}t = 50.4$ (lime solid line). The inner two peaks originate from the initial bi-stream electrons shifting toward lower velocities due to the release of their free energies, while the outer two peaks appear at velocities larger than the initial drift velocity (i.e., $> U_{de, 0}/c = \pm 0.47$). 
The formation of these outer two peaks could be attributed to the significantly larger and higher-amplitude $\phi$ wells (especially after $\omega_{norm}t = 42$) in the strongly magnetized plasma  (see row [$b2$] and panel [$e2$] in Fig.~\ref{Fig1}). These larger and stronger $\phi$ wells imply more extended and stronger localized parallel electrostatic fields (see the fifth panel in row [$c2$] of Fig.~\ref{Fig1}), which form broader and steeper electrostatic wave fronts, thereby accelerating a greater fraction of electrons near the well edges (i.e., via wave-surfing or surfatron acceleration~\citealp{Katsouleas&Dawson_1983PhRvL..51..392K, Dieckmann_etal_2004ApJ...617.1361D, Osmane&Hamza_2014NPGeo..21..115O}). 
Subsequently, these four peaks collectively provide free energy for further excitation of waves in the strongly magnetized plasma. Eventually, as the free energy is released, the $f(U_x)$ gradually evolves into a plateau (i.e., $\partial f(U_x)/ \partial U_x \approx 0$) around $U_x = 0$ in both plasmas.

In contrast, in the weakly magnetized plasma, the $\phi$ wells are small and weak, and the parallel electrostatic fields are correspondingly confined and weak (see rows [$b1$], [$c1$], and panel [$e1$] of Fig.~\ref{Fig1}). Consequently, $f(U_x)$ in the weakly magnetized plasma just smoothly thermalizes without forming multiple peaks (panel [$b1$] of Fig.~\ref{Fig6}).
Features of $f(U_x)$ in the differently magnetized plasmas correspond well to the corresponding phase-space distributions $f(x,U_x)$ shown in rows [$d1$] and [$d2$] of Fig.~\ref{Fig1}. These $f(x,U_x)$ distributions exhibit prominent phase-space vortices and separatrix bifurcations in the the strongly magnetized plasma, whereas in the weakly magnetized plasma, the phase-space bifurcation is suppressed and most electrons have velocities less than $U_{de, 0}$.

In total, at the end of the system evolution (i.e., $\omega_{norm}t = 159.5$), the electrons exhibit distinct anisotropic heating in the two plasmas. In the weakly magnetized plasma, the electrons experience almost isotropic heating (see the inset of panel [$b1$] of Fig.~\ref{Fig6}). In contrast, in the strongly magnetized plasma, the electron heating primarily occurs in the parallel direction (see the inset of panel [$b2$] of Fig.~\ref{Fig6}). 
Although the parallel heating in the strongly magnetized plasma is moderately stronger than that in the weakly magnetized one, the latter includes substantial perpendicular heating as well. Consequently, the overall electron heating in the weakly magnetized plasma is much stronger than that in the strongly magnetized plasma. This implies that, in the weakly magnetized plasma, a large fraction of the free energy carried by the electron bi-streams is ultimately converted into the thermal energy of the electrons themselves, leaving less energy for the waves. Therefore, this perfectly corresponds to the fact that the energy carried by the excited waves is higher in the strongly magnetized plasma than in the weakly magnetized one.

\subsection{Conclusions}
\label{conclusions}

In this study, we have systematically investigated the formation and evolution of the kinetic phase-space structures (i.e., the $\phi$ wells and the EHs), the subsequent wave excitations associated with the $\phi$ wells, as well as the electron acceleration and heating in charge- and current-neutralized plasmas of moderately relativistic electron bi-streams under weak and strong magnetizations via 2.5-dimensional (2.5D) particle-in-cell simulations. The main conclusions are summarized as follows.

First, in both weakly and strongly magnetized plasmas, the electron/electron bi-stream instability excites electrostatic beam-mode waves initially. The subsequent nonlinear evolution of these waves leads to the formation of alternating positive and negative (electrostatic potential) $\phi$ wells, as well as electron holes (EHs). The $\phi$ wells, however, form earlier than the EHs. Both the EHs and the subsequent excitations of electrostatic and electromagnetic waves are fundamentally rooted in the $\phi$ wells.
Regarding the excitation of electrostatic and electromagnetic waves, the localized electrostatic fields ($\vec{E}_{es} = - \nabla \phi$) within the $\phi$ wells drive electrons to undergo bounce oscillations and perpendicular $\vec{E}_{\perp} \times \vec{B}_0$ drifts. These motions, together with other localized current components (e.g., inductive and diamagnetic currents), constitute the localized current sources responsible for wave excitation, consistent with the prediction of the Poynting theorem ($P = -\mathrm{Re}[\tilde{\vec{J}}^* \cdot \tilde{\vec{E}}] > 0$). It should be noted that despite the intense localized currents and wave excitations, the plasma as a whole remains macroscopically charge-neutral and current-free. Overall, the energy carried by the excited electrostatic waves is approximately an order of magnitude higher than that of the excited electromagnetic waves in both plasmas.

Second, the magnetized background fundamentally alters the current dynamics and wave excitation. In the weakly magnetized plasma, the perpendicular (e.g., $\vec{E}_{\perp} \times \vec{B}_0$) drift induces a prominent symmetry breaking in perpendicular velocity space, characterized by a positive gradient ($\partial f/\partial U_\perp > 0$) and a significantly enhanced perpendicular current. This non-Maxwellian feature acts as a free-energy source, exciting the perpendicularly propagating electrostatic Bernstein and electromagnetic electron cyclotron harmonics~\citep[e.g.,][]{Brinca_etal_2003JGRA..108.1038B, Muschietti_etal_2013JGRA..118.2267M, Janhunen_etal_2018PhPl...25h2308J, Benacek&Karlicky2019ApJ...881...21B, Zhou_etal_2020ApJ...891...92Z}. 
In non-perpendicular directions, besides the most-energized electrostatic beam-mode waves associated with the generation of the $\phi$ wells and EHs, the electrostatic Langmuir waves and the electromagnetic whistler, Z-, O-, X-mode waves are excited by the localized currents driven by the $\phi$ wells. In addition, wave-wave interactions also contribute to the excitation of the X-mode waves during the nonlinear quasi-equilibrium stage of the electron bi-stream system in the weakly magnetized plasma.

In contrast, the strong magnetic field suppresses the perpendicular drifts in the strongly magnetized plasma, leaving the perpendicular velocity distribution nearly Maxwellian and thereby suppressing both the strong perpendicular currents and the excitation of these perpendicularly propagating (Bernstein and electron cyclotron) harmonics. Consequently, wave excitation in the strongly magnetized plasma is predominantly driven by the parallel current ($J_x$). In turn, the excited electrostatic (Langmuir) waves propagate mainly in the parallel direction, while the electromagnetic (whistler, Z-, and O-mode) wave energies are primarily distributed at large propagation angles $\theta$ (i.e., oblique propagation). Note that the X-mode excitation is negligible in the strongly magnetized plasma.

Overall, both the excited electrostatic and electromagnetic waves each carry about an order of magnitude more energy in the strongly magnetized plasma than in the weakly magnetized plasma, although the perpendicular dynamics are constrained in the strongly magnetized plasma. The unconstrained perpendicular dynamics in the weakly magnetized plasma perhaps offer additional channels for the dissipation of energy from the $\phi$ wells and EHs; consequently, the lifetimes of the $\phi$ wells and EHs are much shorter than those in the strongly magnetized plasma.

Third, the restricted perpendicular motion also strongly influences the parallel phase-space ($x$--$U_{x}$) dynamics as well as the parallel velocity distribution $f(U_x)$ in the strongly magnetized plasma. The $\phi$ wells in the strongly magnetized plasma are generally wider in area (especially elongated in the perpendicular direction, i.e., almost 1D structures) and larger in amplitude, sustaining stronger localized parallel electrostatic fields ($E_\parallel = -\partial \phi / \partial x$) over a larger spatial extent near the $\phi$ well edges. The large amplitude and broad spatial extent of the $E_\parallel$ fields imply steeper and broader electrostatic wave fronts, thereby allowing a greater fraction of electrons near the $\phi$ well edges to be accelerated to higher velocities via wave-surfing acceleration. These accelerated electrons manifest as both bifurcations in the ($x$--$U_{x}$) space and a distinct four-peak structure in $f(U_x)$: the inner peaks are formed by the thermalization of the initial bi-stream electrons, while the outer peaks (with velocities exceeding the initial drift velocity), together with the phase-space bifurcations, are formed by these highly accelerated electrons. In contrast, the $\phi$ wells in the weakly magnetized plasma are fragmented, smaller in area and weaker in amplitude, which lead to a smooth thermalization of $f(U_x)$ only, without phase-space bifurcations.

Finally, after the free-energy release of the bi-stream electrons, the electrons exhibit distinct anisotropic heating in the two differently magnetized plasmas. Similar to the localized current, the dominant electrostatic wave excitation, as well as the evolution of the velocity and phase-space distributions in the strongly magnetized plasma, where changes are predominantly confined to the parallel direction, the electron heating there also occurs mainly in the direction parallel to the magnetic field due to the restricted perpendicular degrees of freedom. In contrast, the electron heating in the weakly magnetized plasma is essentially isotropic, corresponding to the mostly isotropic localized current, as well as the electrostatic and electromagnetic wave excitations over a broader range of propagation angles $\theta$ there.

\subsection{Discussion}
\label{discussion}

The findings in this study not only reveal the fundamental kinetic processes in magnetized bi-stream plasmas but also offer new insights into several unresolved astrophysical phenomena and open questions for future research.

First, the perpendicularly propagating electrostatic Bernstein and electromagnetic electron cyclotron harmonics, which are excited by the localized perpendicular currents associated with the $\phi$ wells in the weakly magnetized plasma, may provide a novel generation mechanism for zebra-pattern solar radio bursts. 
Furthermore, the excitations of the electromagnetic O- and X-mode waves in differently magnetized electron bi-stream plasmas differ significantly from those in classical coherent emission mechanisms, i.e., the plasma emission mechanism (which relies on the nonlinear conversion of Langmuir waves) and the electron cyclotron maser (ECM) instability (which requires specific velocity distributions with $\partial f/\partial U_\perp > 0$).
The electromagnetic O- and X-mode waves here are directly excited by the localized currents associated with the $\phi$ wells. This presents an alternative, current-driven mechanism for generating electromagnetic radiation in plasmas.

Notably, $\partial f/\partial U_\perp > 0$ is required not only by the ECM instability but also by the classical mechanisms of zebra-pattern solar radio bursts (e.g., the double plasma resonance model). The formation mechanism of $\partial f/\partial U_\perp > 0$ remains debated, with various hypotheses proposed so far. This study, however, demonstrates that $\partial f/\partial U_\perp > 0$ can be self-consistently generated by and coincides with sufficiently large perpendicular currents.

Second, our results expand upon classical wave-surfing acceleration theories, e.g.,~\citep{Katsouleas&Dawson_1983PhRvL..51..392K, Dieckmann_etal_2004ApJ...617.1361D, Osmane&Hamza_2014NPGeo..21..115O}. While traditional models typically require an external transverse magnetic field or a non-uniform background magnetic field to confine electrons to the wave front, we demonstrate that the large-amplitude and and broad-spatial-extented $\phi$ wells in strongly magnetized plasmas themselves can act as a strong spatial confining structure. This allows efficient wave-surfing acceleration of electrons even along parallel-propagating electrostatic waves, a scenario previously considered less favorable.

Third, while our study focuses on the electron/electron bi-stream instability, it is well established that $\phi$ wells and EHs can be generated by a variety of systems driven by electron beam instabilities (e.g., electron/ion two-stream or Buneman instability, and the electron bump-on-tail instability). A natural question arises: is the contribution of $\phi$ wells to wave excitation equally important in these different instability regimes, and how do the characteristics of the excited waves differ? This remains an open question for future comparative studies.

Finally, the present simulations are 2.5-dimensional (2.5D). In a fully 3D reality, we expect the excitation of the perpendicularly propagating Bernstein and electron cyclotron harmonics to be even stronger. This is because the perpendicular electric field in 3D would include contributions from both $y$ and $z$ directions: $\vec{E}_{\perp} = -(\partial \phi / \partial y + \partial \phi / \partial z)$. In the 2.5D simulations, the $z$-gradient is assumed to be zero ($\partial \phi / \partial z = 0$), which likely underestimates the perpendicular current and the resulting wave excitation. Future 3D PIC simulations will be necessary to quantify this effect.

\begin{acknowledgments}
The present research at PMO was supported by the Strategic Priority Research Program of the
Chinese Academy of Sciences under grant No.XDB0560000, the project of National Natural Science
Foundation of China No.12003073, 42174195, 11873018, and 11790302. And Jan Ben\'{a}\v{c}ek
acknowledges the support by the German Science Foundation (DFG) project BE 7886/2-1.
We gratefully acknowledge the developers of the ACRONYM code, the Verein zur F\"{o}rderung kinetischer Plasmasimulationen e.V.,
as we as the computing resources in the high-performance computing cluster "Zimo" from the Purple Mountain Observatory of the Chinese Academy of Sciences, Nanjing, China. \end{acknowledgments}

\bibliography{Reference}{}
\bibliographystyle{aasjournalv7}



\end{document}